# Two-dimensional superconducting diode effect in topological insulator/superconductor heterostructure

Soma Nagahama[1], Yuki Sato[2], Minoru Kawamura[2], Ilya Belopolski[2], Ryutaro Yoshimi[3], Atsushi Tsukazaki[1,4], Naoya Kanazawa[5], Kei S Takahashi[2], Masashi Kawasaki[1,2], and Yoshinori Tokura[1,2,6]

[1]Department of Applied Physics and Quantum-Phase Electronics Center (QPEC), The University of Tokyo, Tokyo 113-8656, Japan

[2]RIKEN Center for Emergent Matter Science (CEMS), Wako 351-0198, Japan

[3]Department of Advanced Materials Science, The University of Tokyo, Kashiwa 277-8561, Japan

[4]Institute for Materials Research (IMR), Tohoku University, Sendai 980-8577, Japan

[5]Institute of Industrial Science, The University of Tokyo, Tokyo 153-8505, Japan and

[6]Tokyo College, The University of Tokyo, Tokyo 113-8656, Japan

**Abstract**

The superconducting diode effect (SDE) is characterized by the nonreciprocity of Cooper-pair motion with respect to current direction. In three-dimensional (3D) materials, SDE results in a critical current that varies with direction, making the effect distinctly observable: the material exhibits superconductivity in one direction while behaving as a resistive metal in the opposite direction. However, in genuinely two-dimensional (2D) materials, the critical current density is theoretically zero, leaving the manifestation of SDE in the 2D limit an intriguing challenge. Here, we present the observation of SDE in a heterostructure composed of the topological insulator $Bi_2Te_3$ and the iron-based superconductor Fe(Se,Te)—a candidate for topological superconductor—where superconductivity is confined to the 2D limit. The observed *I-V* characteristics reveal nonreciprocity in the vortex-creep regime, where finite voltages arise due to the two-dimensional nature of superconductivity. Furthermore, our 2D film demonstrates abrupt voltage jumps, influenced by both the current flow direction and the transverse magnetic field direction. This behavior resembles that of 3D materials but, in this case, is driven by the vortex-flow instability, as illustrated by voltage-controlled S-shaped *I-V* curves. These results underscore the pivotal role of vortex dynamics in SDE and provide new insights into the interplay between symmetry breaking and two-dimensionality in topological insulator/superconductor systems.

## Introduction

In conventional superconductors, Cooper pairs form spin singlet states with even parity. However, a mixing of spin triplet component can occur in superconductors with broken inversion symmetry [1], giving rise to nonreciprocal transport properties [2-4]. One of the most striking manifestations of the nonreciprocal phenomena in superconducting systems is the superconducting diode effect (SDE) [5], in which the motion of Cooper pairs depends on the direction of current, allowing dissipationless transport in one direction and dissipative transport in the other (Fig. 1(a)). SDE has been studied in a diverse range of superconducting systems, including metal superlattices [6], ferromagnet/superconductor bilayer [7, 8], flakes of transition metal dichalcogenides [9, 10], kagome superconductor [11], and twisted bilayer or trilayer graphene [12, 13]. In systems where both time and inversion symmetries are broken, the nonreciprocal behaviors, including SDE, can be attributed to magnetochiral anisotropy, expressed as $\boldsymbol{z} \cdot (\boldsymbol{I} \times \boldsymbol{B})$ [14-17], where $\boldsymbol{z}$ is the polar vector indicating the direction of inversion symmetry breaking, $\boldsymbol{I}$ is the current, and $\boldsymbol{B}$ is the magnetic field or the magnetization (Fig. 1(a)). While SDE generally requires both inversion and time-reversal symmetry breakings, a few exceptions have been reported in systems where time-reversal symmetry remains intact [12, 13, 18, 19].

Heterostructure interfaces involving superconducting materials provide an platform for investigating the inversion symmetry broken superconductors [5]. In particular, the interface between a topological insulator and a superconductor has emerged as a compelling materials platform, because the topological surface state (TSS) of the topological insulator inherently breaks the inversion symmetry. Theoretically, it has been predicted that the superconducting state with triplet Cooper pairs can emerge when a superconducting gap is induced on the TSS by the proximity effect [20]. Several experimental studies have reported nonreciprocal resistance in the topological insulator/superconductor heterostructures near the superconducting transition [21-23]. In a previous study [23], we observed nonreciprocal resistance in such heterostructures and found that the degree of nonreciprocity enhances as the superconducting layer is thinned toward the two-dimensional limit. However, these studies have primarily focused on resistance in the low-current regime at temperatures just above the zero-resistance temperature and have not systematically addressed SDE or the nonreciprocity in the critical current regime.

While heterostructure interfaces are well-suited for studying SDE due to the broken inversion symmetry, SDE in the 2D limit is a nuanced issue for the following reason. In a 2D system, the superconducting transition is governed by the Berezinskii-Kosterlitz-Thouless (BKT) transition mechanism [24-26], in which thermally excited vortex and anti-vortex form pairs and do not move freely below the BKT transition temperature $T_{\mathrm{BKT}}$. When a bias current is applied, these pairs can dissociate, and the resulting free vortices undergo creep motion, generating a non-zero voltage. This voltage increases gradually with current, indicating the absence of a well-defined critical current, *i.e.*, a sharp boundary between zero and nonzero voltages. Consequently, defining SDE in 2D superconductors becomes problematic as discussed in a recent theoretical study [27], and the relationship between SDE and two-dimensionality remains unexplored. However, even in the case of 2D superconductors, significant voltage jumps can occur due to instability of vortex dynamics [28, 29], offering an opportunity to investigate SDE-like phenomenon in 2D superconductors with mixing of spin-singlet and triplet components.

In this paper, we report on the observation of SDE in a 2D superconducting system. SDE appears at the voltage jumps associated with the vortex-flow instability as well as in the vortex creep regime. The 2D superconducting system is realized using heterostructure films of topological insulator $Bi_2Te_3$ (BT) and superconductor $FeSe_{0.1}Te_{0.9}$ (FST), as illustrated in Fig. 1(b). The heterostructure can be modelled as a parallel connection of two superconducting components: TSS with a superconducting gap and the superconducting FST layer with broken inversion symmetry due to an electron transfer from BT to FST [30]. While FST possesses a centrosymmetric crystal structure, enhancement of nonreciprocal transport signals was observed by thinning the FST layer, as discussed in the previous work [23]. The enhancement can result from a Rashba band splitting of FST due to the strong structural asymmetry induced by the stacking configuration and the resulting charge transfer. In the present study, we selected the thicknesses of BT and FST to be 10 nm and 2 nm, respectively, to confine the superconductivity in the 2D limit and achieve a sizable SDE.

**Synthesis of topological insulator/superconductor heterostructures**

The films were grown on CdTe(001) substrates using molecular beam epitaxy method. $FeSe_xTe_{1-x}$

exhibits superconductivity over a broad range of Se composition $x$, and we selected $x = 0.1$, where the zero-resistivity temperature reaches its maximum [31]. The successful growth of the BT/FST heterostructures was confirmed by x-ray diffraction, atomic force microscopy (Supplemental Material Fig. S1), and scanning transmission electron microscopy (STEM) (Fig. 1(c)). The cross-sectional STEM image clearly reveals atomically flat interfaces with the FST layer thickness of 4 unit cells (approximately 2 nm). The films were patterned into Hall bars with the lateral dimensions of 100 μm × 100 μm.

**Two-dimensionality of superconductivity**

Due to the reduced thickness comparable to the coherence length (1 nm) [32], the film exhibits 2D superconductivity behavior. Figure 1(d) shows the temperature $T$ dependence of resistance $R$ of the film. The onset of the superconducting transition temperature is around 12.5 K, which is typical of $FeSe_{0.1}Te_{0.9}$ grown on CdTe substrates [31]. The $R$-$T$ curve is well described by the Halperin-Nelson formula $R = R_0 \exp\left(-2b\left(\frac{T_0-T}{T-T_{BKT}}\right)^{1/2}\right)$ with fitting parameters $R_0$, $b$, $T_0$ and $T_{\mathrm{BKT}}$, which characterizes the resistance behavior near the BKT transition in 2D superconductors [33]. From the fit, $T_{\mathrm{BKT}}$ is estimated at 8.0 K. Figure 1(e) shows a log-log plot of current-voltage ($I$-$V$) characteristics measured at various temperatures. The $I$-$V$ curves exhibit power-law behavior, $V \sim I^p$, over a wide temperature range, which is characteristic of 2D superconductors [26]. The exponent $p$ reaches 3 near $T_{\mathrm{BKT}} = 8.0$ K (Supplementary Discussion 1 and Fig. S1) as expected for the BKT transition. To further confirm the 2D nature of the system, we measured the magnetic field angle dependence of the upper critical field $B_{c2}$, as shown in Fig. 1(f). A clear cusp appears at the angle where the magnetic field aligns with the film plane, consistent with the behavior expected for 2D superconductors. The angle dependence shows good agreement with Tinkham model of 2D superconductivity [34] rather than with a model describing 3D superconductivity (Supplementary Discussion 2 and Fig. S2).

**Nonreciprocal *I-V* characteristics under magnetic fields**

Figures 2(a)-2(c) show the $|I|$-$|V|$ characteristics at a temperature $T = 2.2$ K, well below $T_{BKT}$, under magnetic fields of $B$ = +3 T, -3 T, and 0 T, respectively. The magnetic fields are applied parallel to the film plane along the *y*-axis direction (perpendicular to the current direction), as illustrated in the inset of Fig. 2(a). As current $|I|$ increases, the voltage abruptly jumps at approximately $|I|$ = 2.9 mA. When decreasing current, the voltage drops back to the noise floor around 2.0 mA, forming a hysteresis loop. Such hysteretic *I-V* curves accompanied by voltage jumps can be observed in superconducting thin film or nanowire due to the instability of vortex dynamics, local heating effect, or quantum phase slip [28, 35-38]. In the following discussion, we focus on the current at which the voltage jump occurs and refer to it as $I^*$.

The most remarkable feature in Figs. 2(a)-2(c) is that the values of $|I^*|$, where the voltage exhibits a distinct jump, depend on the current directions as well as on the magnetic field directions. Specifically, $|I^*_+|$ is larger than $|I^*_-|$ at $B$ = +3 T in Fig. 2(a), while the opposite trend is observed at $B$ = -3 T in Fig. 2(b), where $|I^*_+|$ and $|I^*_-|$ denote the absolute values of the current at the voltage jumps for positive and negative current directions, respectively. These results clearly demonstrate the SDE in the two-dimensional superconductor. The magnitude of SDE, defined at the voltage jumps as $\eta = \frac{|I_+^*|-|I_-^*|}{|I_+^*|+|I_-^*|}$ is found to be 1.5 %. The value of $\eta$ is much smaller than those reported in literature [9,10,12]. We speculate that the difference may arise from the way of inversion symmetry breaking. Specifically, inversion symmetry in our system is broken by sandwiching the centrosymmetric material (FST) with $Bi_2Te_3$ and CdTe while that is broken by the crystal symmetry in the materials reported in the literature. We confirmed that no SDE is observed at $|I^*|$ when $B = 0$ (Fig. 2(c)). Additionally, the current value at which the voltage drops back to zero while decreasing $|I|$ is reciprocal regardless of the magnetic field.

Examining the $|I|$-$|V|$ curves on logarithmic scale (Figs. 2(d) and 2(e)) reveals clear asymmetry between the positive (red) and negative (blue) current directions in the low-voltage regime prior to the voltage jumps. In the case of $B = 0$, the $|I|$-$|V|$ curves for both current directions overlap, exhibiting a reciprocal behavior (Fig. 2(f)). The appearance of small but finite voltages prior to the voltage jumps is characteristic of 2D superconductivity and is attributed to the creep of the dissociated superconducting vortices [26]. The fact that the nonreciprocal feature appears in this low-voltage regime under the magnetic field suggests that vortex dynamics plays a crucial role in the observed nonreciprocity. Furthermore, the reversal of the nonreciprocity by the reversal of the in-plane magnetic field direction is consistent with the theoretical framework of magnetochiral anisotropy [14].

Taking advantage of SDE observed at the voltage jumps, we demonstrate voltage rectification at $B$= +3 T, as shown in Fig. 2(g). The current was cyclically changed between +2.85 mA -2.85 mA through 0 mA with a period of 12 s, where $|I|$ = 2.85 mA lies between $|I^*_+|$ and $|I^*_-|$. Sizable voltages are obtained only when $I < 0$, consistent with previous demonstrations of voltage rectification using SDE [10, 12, 18].

**Magnetic field and temperature dependence**

We investigate the magnetic field and temperature dependence of SDE (See End Matter for the temperature dependence). Figure 3(a) shows the magnetic field dependence of the difference in $|I^*|$ defined as $\Delta I^* \equiv (|I_+^*| - |I_-^*|)/2$. $\Delta I^*$ develops linearly with the magnetic field and begins to saturate around 3 T. The magnetic field dependence looks qualitatively similar to that of the nonreciprocal resistance (NR) observed in the second harmonic voltage at temperatures above $T_{BKT}$ in the previous study [23] (Supplementary Discussion 3 and Fig. S3). This similarity implies that the in-plane magnetic field could exert a similar influence on moving vortices in both cases.

We have also investigated the magnetic field angle dependence by rotating the magnetic fields in both *yz*- and *xy*-planes at $|B| = 3$ T (Figs. 3(b)-(c)). In the field angle scan in the *yz*-plane, $\Delta I^*$ is largest when the magnetic field is along the *y*-axis (in-plane direction). This observation rules out the possibility of SDE arising from the extrinsic Meissner screening effect [39], because such an effect would be pronounced when the magnetic field is aligned in the out-of-plane direction. The antisymmetric angle dependence with respect to $\theta = 180°$ also excludes the possibility of the nonreciprocity originating from the device structure or the contact resistance, as discussed in earlier studies [40,41]. When the magnetic field is rotated in the *xy*-plane, $\Delta I^*$ nearly follows the sinusoidal curve, reflecting the dominant role of the magnetochiral anisotropy.

**Voltage jumps and vortex dynamics**

Finally, we discuss the physical implications of the voltage jumps in the $|I|$-$|V|$ curves. By conducting two different measurement methods of current scanning ($I$-scan, Fig. 4(a)) and voltage scanning ($V$-scan, Fig. 4(b)) under the in-plane magnetic field of $B$ = +3 T, two different $|I|$-$|V|$ curves were obtained as shown in Figs. 4(c) and 4(d). The $V$-scan curves exhibit S-shaped forms, where the extrema of the current correspond to the current $I^*$ at which the voltage jumps occur in the $I$-scan curves. Notably, in the present 2D superconducting system, $I^*$ does not represent the Cooper-pair breaking current density which would define the boundary between zero and finite resistance. Indeed, assuming that the supercurrent flows homogeneously in the 2-nm-thick FST layer, the current density corresponding to $|I^*|$ is approximately $1.5 \times 10^6$ A/cm$^2$, which is significantly smaller than the estimated Cooper-pair breaking current density (~ $10^7$ A/cm$^2$) for FST [42]. This discrepancy suggests that the observed voltage jumps have an origin other than the Cooper-pair breaking.

The S-shaped $I$-$V$ curves are reminiscent of nonlinear conductivity arising from the non-equilibrium distribution of quasi-particles, as previously studied for the mixed state of type-II superconductors [43,44]. As the vortex velocity increases, the number of quasi-particle excitations decreases within the vortex core, leading to a decrease in the viscous damping of the vortex motion [45]. Consequently, when the velocity reaches a critical value, the current starts to diminish, resulting in the characteristic S-shaped $I$-$V$ curve. Thus, the observed voltage jumps at $I^*$ are interpreted as a result of vortex-flow instability that occurs when the vortex velocity exceeds the critical value.

In the $V$-scan measurements, nonreciprocity appears as the asymmetry in the peak heights of the current ($I$*) as well as in the peak position with respect to the voltage ($V$*). Figure 4(e) presents a magnified view of the small-voltage region, revealing clear directional asymmetry in the voltages $V$* at which the current reaches the extremum. The asymmetry in $V^*$, $\Delta V^* \equiv (|V_+^*| - |V_-^*|)/2$, which is negligibly small when the time reversal symmetry is preserved (zero magnetic field), while it develops abruptly when the time reversal symmetry is broken (under the magnetic fields $B > 0$), as shown in Fig. 4(f). $\Delta V^*/V_{\mathrm{av}}^*$ is negative, - 5 % or more, indicating that $|V_+^*| < |V_-^*|$ under the magnetic fields $B > 0$. Because the voltage $V$, generated by vortex motion, is in proportion to the product of the vortex density $n$ and the vortex velocity $v$, this asymmetry indicates that the small nonreciprocal modulation of vortex velocity $v$ and vortex density $n$ due to the in-plane magnetic fields [3], highlighting the nonreciprocal nature of vortex motion.

When the in-plane magnetic field is varied, the average value of $V_{\mathrm{av}}^*$, defined as $V_{\mathrm{av}}^* \equiv (|V_+^*| + |V_-^*|)/2$, remains nearly unchanged (Fig. 4(g)). The nearly $B$-independent $V_{\mathrm{av}}^*$ indicates that the vortex density $n$ is not largely changed by the in-plane magnetic fields, being consistent with the two-dimensional nature of the present superconducting system. This contrasts to the case of out-of-plane magnetic fields in which $V_{\mathrm{av}}^*$ shows a linear increase, reflecting increase in the number of field induced vortices (Supplementary Discussion 4 and Fig. S4). These observations suggest that thermally excited out-of-plane vortices, rather than field-induced vortices, are responsible for the SDE, contrasting with the proposed mechanism that the SDE is associated with to the depinning of the in-plane-field-induced vortices [42].

Our observations highlight the crucial role of vortex dynamics in the nonreciprocal superconducting transport and provide new insights into the physics of pairing mechanism in inversion-symmetry-broken superconducting systems.

**References**

[1] V. Kozii *et al.*, Odd-parity superconductivity in the vicinity of inversion symmetry breaking in spin-orbit-coupled systems, Phys. Rev. Lett. **115**, 207002 (2015).

[2] R. Wakatsuki, and N. Nagaosa, Nonreciprocal current in noncentrosymmetric Rashba superconductors, Phys. Rev. Lett. **121**, 026601 (2018).

[3] S. Hoshino *et al.*, Nonreciprocal charge transport in two-dimensional noncentrosymmetric superconductors, Phys. Rev. B **98**, 054510 (2018).

[4] J. Zhai *et al.*, Nonreciprocal transport in $KTaO_3$-based interface superconductors with parity mixing, Phys. Rev. Lett. **134**, 236303 (2025).

[5] M. Nadeem, M. S. Fuhrer, and X. Wang, The superconducting diode effect, Nat. Rev. Phys. **5**, 558 (2023).

[6] F. Ando *et al.*, Y. Observation of superconducting diode effect, Nature **584**, 373 (2020).

[7] A. Gutfreund *et al.*, Direct observation of a superconducting vortex diode, Nat. Commun. **14**, 1630 (2023).

[8] R. Cai, I. Žutić, and W. Han, Superconductor/ferromagnet heterostructures: A platform for superconducting spintronics and quantum computation, Adv. Quantum Technol. **6**, 2200080 (2023).

[9] L. Bauriedl *et al.*, Supercurrent diode effect and magnetochiral anisotropy in few-layer $NbSe_2$, Nat. Commun. **13**, 4266 (2022).

[10] B. Pal *et al.*, Josephson diode effect from Cooper pair momentum in a topological semimetal, Nat. Phys. **18**, 1228 (2022).

[11] T. Le *et al.*, Superconducting diode effect and interference patterns in kagome $CsV_3Sb_5$, Nature **630**, 64 (2024).

[12] J. X. Lin *et al.*, Zero-field superconducting diode effect in small-twist-angle trilayer graphene, Nat. Phys. **18**, 1221 (2022).

[13] J. X. Hu *et al.*, Josephson diode effect induced by valley polarization in twisted bilayer graphene, Phys. Rev. Lett. **130**, 266003 (2023).

[14] Y. Tokura and N. Nagaosa, Nonreciprocal responses from non-centrosymmetric quantum materials, Nat. Commun. **9**, 3740 (2018).

[15] J. J. He, Y. Tanaka, and N. Nagaosa, A phenomenological theory of superconductor diodes, New J. Phys. **24**, 053014 (2022).

[16] A. Daido, Y. Ikeda, and Y. Yanase, Intrinsic superconducting diode effect, Phys. Rev. Lett. **128**, 037001 (2022).

[17] N. F. Q. Yuan, and L. Fu, Supercurrent diode effect and finite-momentum superconductors, PNAS **119**, e2119548119 (2022).

[18] H. Wu *et al.*, The field-free Josephson diode in a van der Waals heterostructure, Nature **604**, 653 (2022).

[19] F. Liu *et al.*, Superconducting diode effect under time-reversal symmetry, Sci. Adv. **10**, eado1502 (2024).

[20] L. Fu, and C. L. Kane, Superconducting proximity effect and Majorana fermions at the surface of a topological insulator, Phys. Rev. Lett. **100**, 096407 (2008).

[21] K. Yasuda *et al*., Nonreciprocal charge transport at topological insulator/superconductor interface, Nat. Commun. **10**, 2734 (2019).

[22] M. Masuko *et al.*, Nonreciprocal charge transport in topological superconductor candidate $Bi_2Te_3/PdTe_2$ heterostructure, npj Quant. Mater. **7**, 104 (2022).

[23] S. Nagahama *et al*., Control of nonreciprocal charge transport in topological insulator/superconductor heterostructures with Fermi level tuning and superconducting-layer thickness, Phys. Rev. B **112**, L121110 (2025).

[24] M. Tinkham, Introduction to superconductivity, 2nd edition, Dover (2004).

[25] Y. Saito, T. Nojima, and Y. Iwasa, Highly crystalline 2D superconductors, Nat. Rev. Mater. **2**, 16094 (2017).

[26] A. M. Kadin, K. Epstein, and A. M. Goldman, Renormalization and the Kosterlitz-Thouless transition in a two-dimensional superconductor, Phys. Rev. B **27**, 6691 (1983).

[27] N. Nunchot, and Y. Yanase, Nonlinear diode effect and Berezinskii-Kosterlitz-Thouless transition in purely two-dimensional noncentrosymmetric superconductors, Phys. Rev. B **111**, 094515 (2025).

[28] S. K. Ojha *et al*., Flux-flow instability across Berezinskii Kosterlitz Thouless phase transition in $KTaO_3$ (111) based superconductor, Commun.Phys. **6**, 126 (2023).

[29] Y. Saito *et al*., Dynamical vortex phase diagram of two-dimensional superconductivity in gated $MoS_2$, Phys. Rev. Mater. **4**, 074003 (2020).

[30] K. Owada, *et al.* Electronic structure of a $Bi_2Te_3$/FeTe heterostructure: Implications for unconventional superconductivity, Phys. Rev. B **100**, 064518 (2019).

[31] Y. Sato *et al*., Molecular beam epitaxy of superconducting $FeSe_xTe_{1-x}$ thin films interfaced with magnetic topological insulators, Phys. Rev. Mater. **8**, L041801 (2024).

[32] Y. Sato *et al.*, Non-Fermi liquid transport and strong mass enhancement near the nematic quantum critical point in $FeSe_xTe_{1-x}$ thin films, Phys. Rev. B **112**, L041121 (2025).

[33] B. I. Halperin, and D. R. Nelson, Resistive transition in superconducting films, J. Low Temp. Phys. **36**, 599 (1979).

[34] M. Tinkham, Effect of fluxoid quantization on transitions of superconducting films, Phys. Rev. **129**, 2413 (1963).

[35] M. Liang, and M. N. Kunchur, Vortex instability in molybdenum-germanium superconducting films, Phys. Rev. B **82**, 144517 (2010).

[36] W. J. Skocpol, M. R. Beasley and M. Tinkham, Self-heating hotspots in superconducting thin-film microbridges, J. Appl. Phys. **45**, 4054 (1974).

[37] S. Michotte *et al.*, Current–voltage characteristics of Pb and Sn granular superconducting nanowires, Appl. Phys. Lett. **82**, 4119 (2003).

[38] M. Tinkham *et al.*, Hysteretic $I-V$ curves of superconducting nanowires, Phys. Rev. B **68**, 134515. (2003).

[39] Y. Hou *et al.*, Ubiquitous superconducting diode effect in superconductor thin films, Phys. Rev. Lett. **131**, 027001 (2023).

[40] T. Terashima *et al.*, Apparent nonreciprocal transport in FeSe bulk crystals, Phys. Rev. Lett. **111**, 054521 (2025).

[41] U. Nagata *et al.*, Field-free superconducting diode effect in layered superconductor FeSe, Phys. Rev. Lett. **134**, 236703 (2025).

[42] Y. Kobayashi *et al.*, A scaling relation of vortex-induced rectification effects in a superconducting thin-film heterostructure, Commun. Phys. **8**, 196 (2025).

[43] A. I. Larkin, and Y. N. Ovchinnikov, Nonlinear conductivity of superconductors in the mixed state, Sov. Phys. JETP **41**, 960 (1975).

[44] A. I. Bezuglyj, and V. A. Shklovskij, Effect of self-heating on flux flow instability in a superconductor near $T_c$, Physica C **202**, 234 (1992).

[45] W. Klein *et al*., Nonlinearity in the flux-flow behavior of thin-film superconductors, J. Low Temp. Phys **61**, 413 (1985).

[47] S. Nagahama *et al*., Data for "Two-dimensional superconducting diode effect in topological insulator/superconductor heterostructure", https://doi.org/10.5281/zenodo.17655307

**ACKNOWLEDGEMENTS**

This work was supported by JSPS KAKENHI Grants (No.22H04958, No. 24K17020, No. 22K18965, No. 23H04017, No. 23H05431, No. 23H05462, No. 23H04862, 24H00417, 24H01652, and 25H02126), JST FOREST (Grant No. JPMJFR2038), JST CREST (Grant No. JPMJCR1874 and No. JPMJCR23O3), Mitsubishi Foundation, Sumitomo Foundation, Tanaka Kikinzoku Memorial Foundation, the special fund of the RIKEN TRIP initiative (Many-body Electron Systems).

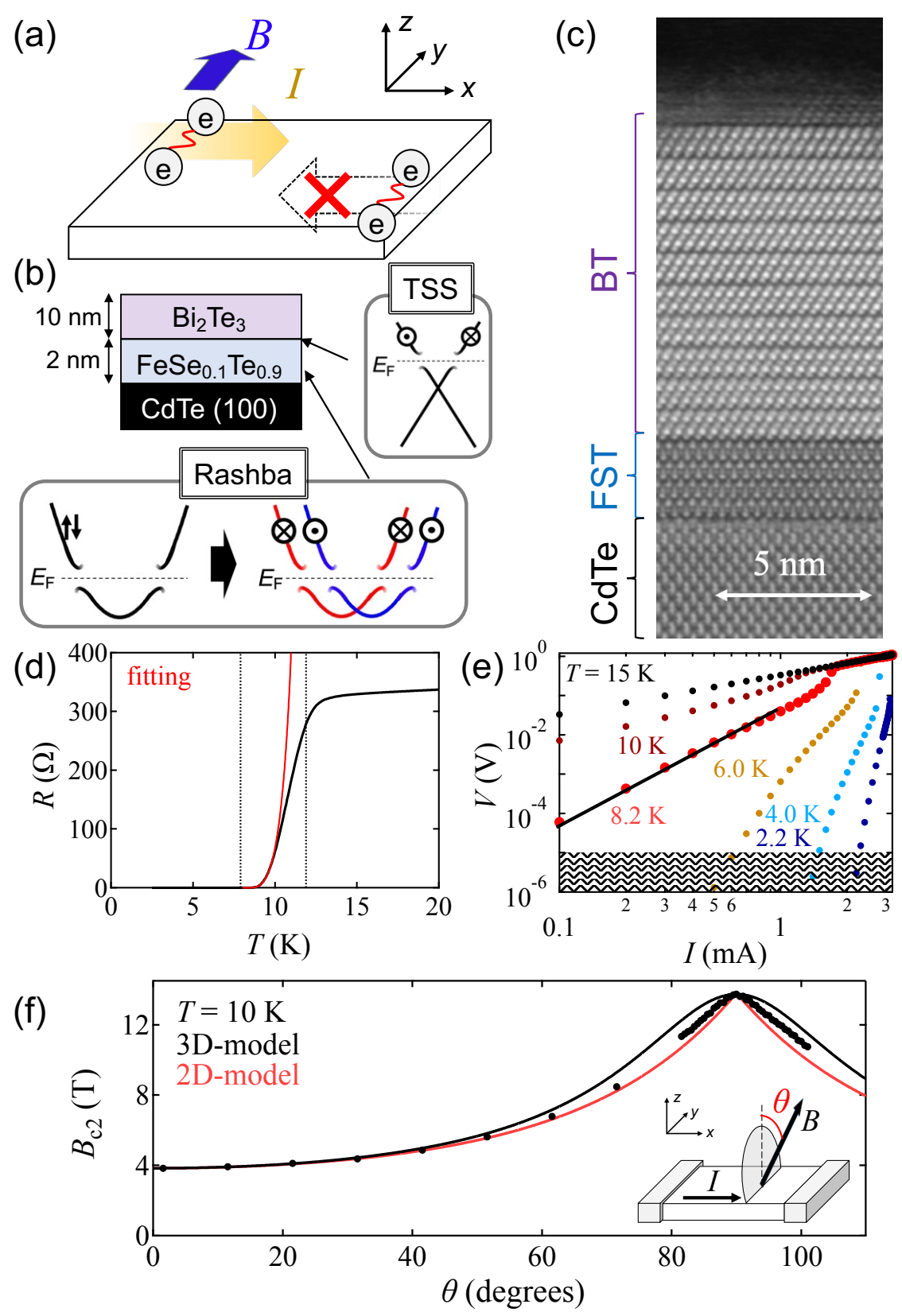

FIG. 1. **Characterizations of a $Bi_2Te_3/FeSe_{0.1}Te_{0.9}$ heterostructure**. (a) Schematic of superconducting diode effect (SDE). (b) Schematics of $Bi_2Te_3/FeSe_{0.1}Te_{0.9}$ heterostructure and band diagrams under inversion symmetry breaking. (c) Cross-sectional scanning transmission electron microscopy image of the heterostructure. (d) Temperature dependence of the resistance of the heterostructure (black). Red curve represents the fit obtained using the Halperin-Nelson formula. The fitting parameters $T_{BKT}$ = 8.0 K and $T_0$ = 11.9 K are indicated by horizontal dashed lines. (e) Log-log plot of $I$-$V$ characteristic curves at $T$ = 2.2, 4.0, 6.0, 8.2, 10.0, and 15.0 K. Black line is a guide to eyes representing $V \sim I^3$. (f) Magnetic field angle dependence of the upper critical field at $T$ = 10 K. Red and black curves represent the theoretical curves expected from the 2D and 3D models, respectively.

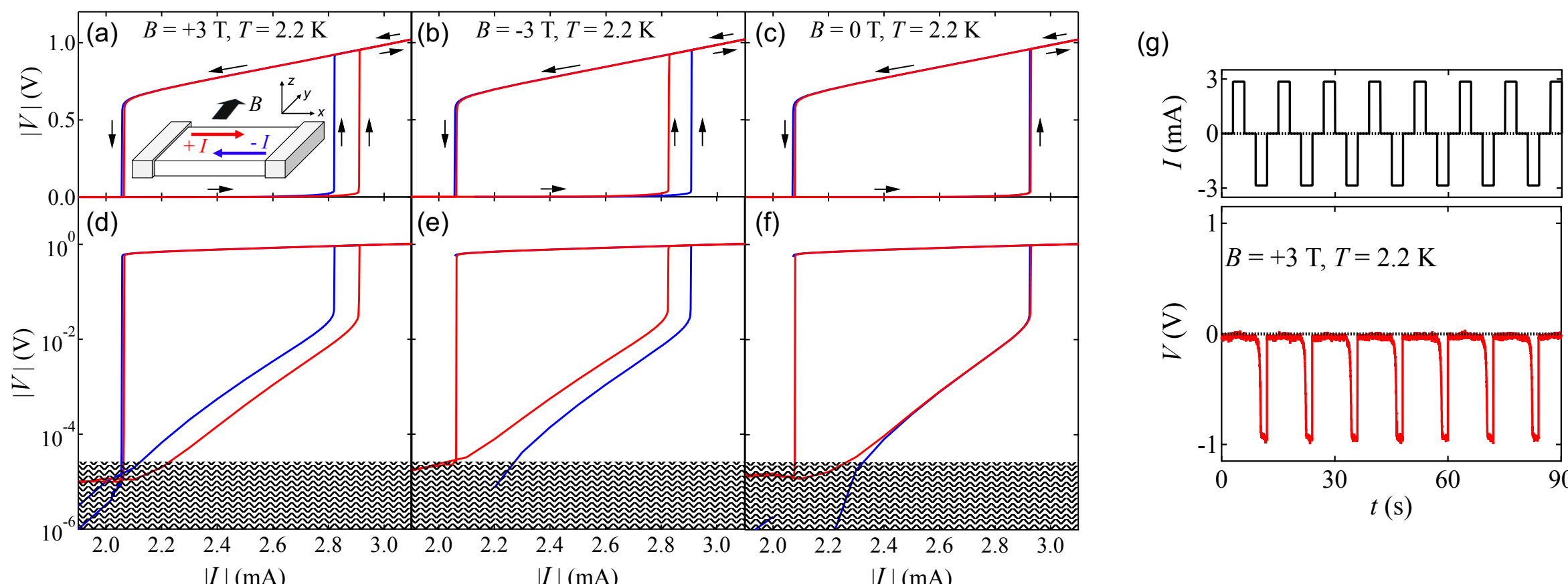

FIG. 2. **Observation of superconducting diode effect (SDE)**. (a)-(c) $|I|$-$|V|$ characteristics at $T = 2.2$ K under $B = +3$, -3, and 0 T, respectively. The curves for $I > 0$ ($I < 0$) are shown in red (blue). Arrows indicate the current sweep directions. (d)-(f) Logarithmic scale plot of the same data as (a)-(c). The shadowed area lies below the noise floor. (g) Demonstration of voltage rectification by using SDE. Temporal profile of the input current and output voltage are shown in the top and bottom panels, respectively. The current is cyclically switched among +2.85, 0, and -2.85 mA with a period of 12 s.

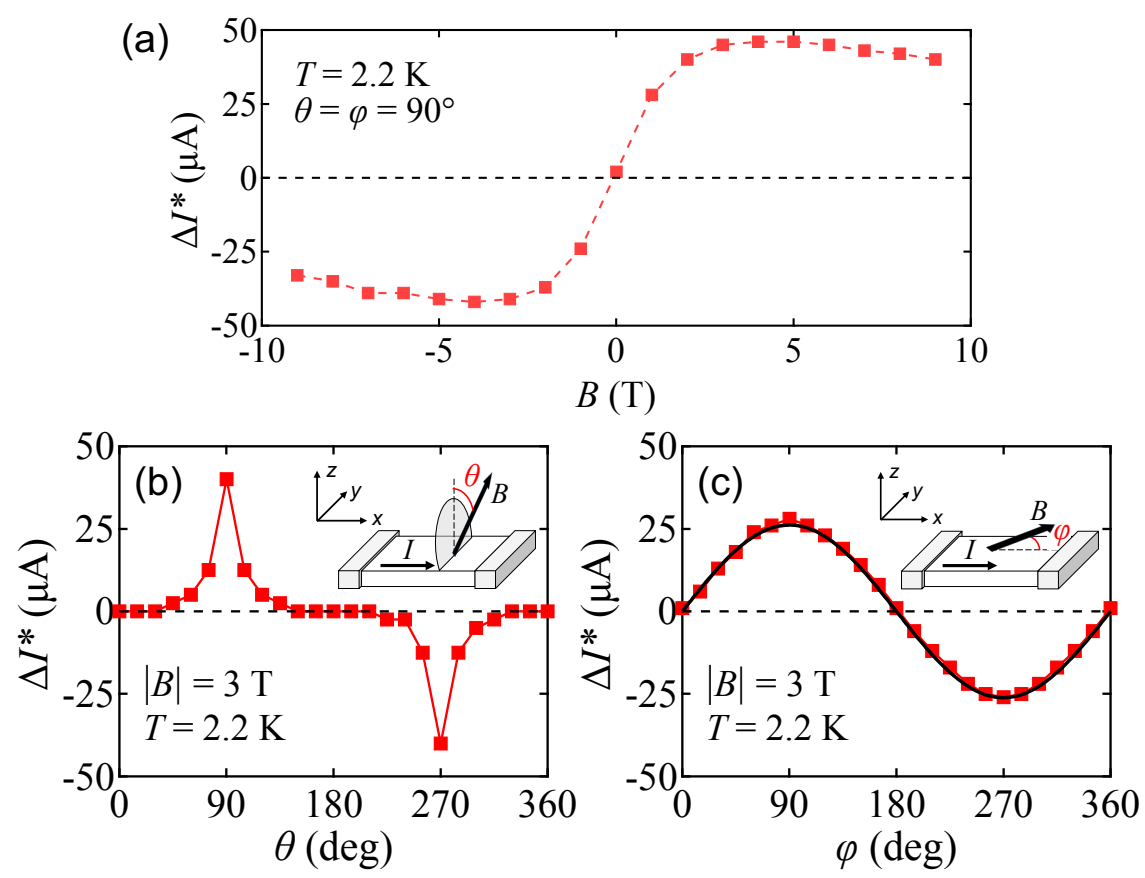

FIG. 3. **Magnetic field and angle dependence of SDE**. (a) Magnetic field dependence of the current difference $\Delta I^* = (|I_+^*| - |I_-^*|)/2$. (b)-(c)**,** Magnetic field angle dependence of $\Delta I^*$. Magnetic field is rotated in the *yz*-plane in (b) and *xy*-plane in (c), as shown in each inset. Different devices were used for each measurement.

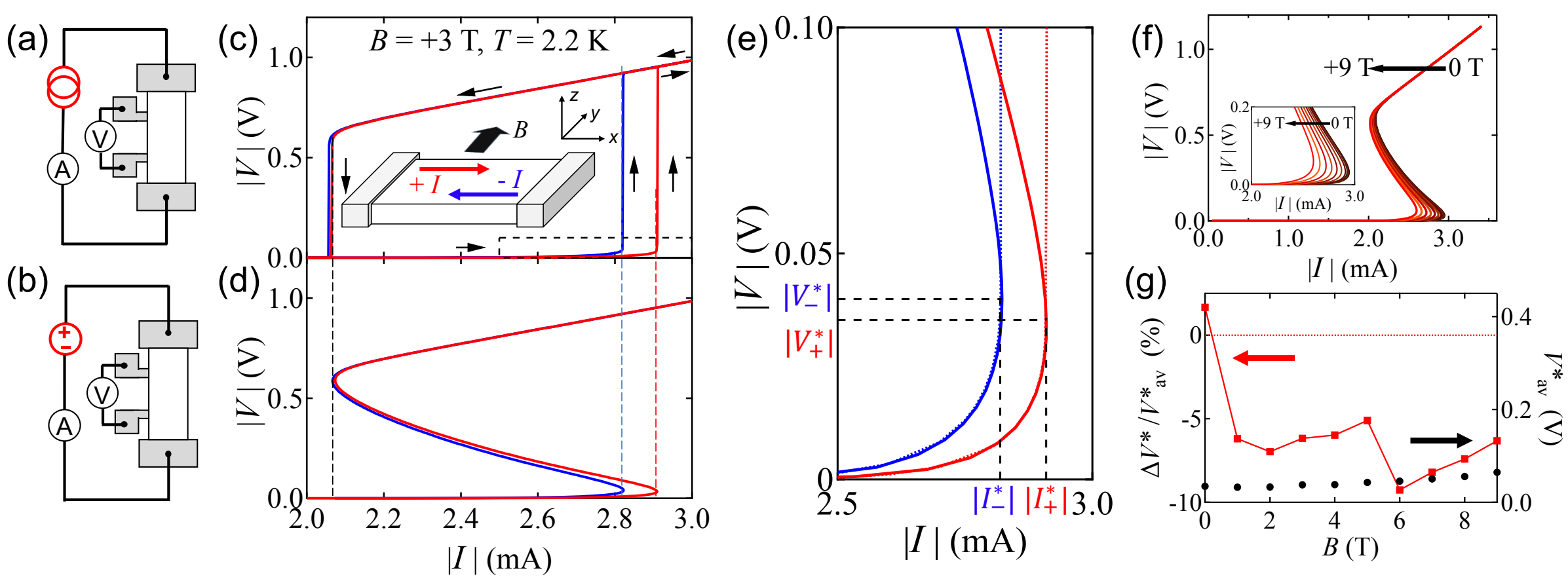

FIG. 4. **Observation of an instability in *I*-*V* curve**. (a)-(b) Schematic circuit diagrams for (a) *I*-scan and (b) *V*-scan measurements. (c) |*I*|-|*V*| characteristics obtained by the *I*-scan measurement at $B = +3$ T and $T = 2.2$ K. (d) |*I*|-|*V*| characteristics obtained by the *V*-scan measurement at $B = +3$ T and $T = 2.2$ K. The dashed lines are guide to eyes. (e) Magnified view of the |*I*|-|*V*| curves. The *I*-scan (dashed) and *V*-scan (solid) curves are overlaid. (f) |*I*|-|*V*| characteristics obtained by *V*-scan measurement under various in-plane magnetic fields from $B = 0$ T to 9 T with an increment of 1 T. The data for $I > 0$ are shown. The inset shows the magnified view of the figure. (g) Magnetic field dependence of $\Delta V^*/V^*_{\mathrm{av}}$ (red) and $V^*_{\mathrm{av}}$ (black).

**End Matter**

**Appendix: Temperature dependence of $|I|$-$|V|$ characteristics**

Figures E1(a) and E1(b) show the evolution of the nonreciprocal $|I|$-$|V|$ characteristics with temperature. As temperature increases, the voltage jump becomes less distinct, and the $|I|$-$|V|$ curves gradually change to ordinary Ohmic behavior. As shown in Fig. E1(b), the power-law exponents of the $|I|$-$|V|$ curves in the low-voltage vortex-creep regime are nearly the same for both positive and negative current directions. Figure E1(c) shows the temperature dependence of $\Delta I^*$ and $I^*_{\rm av} \equiv (|I^*_+| + |I^*_-|)/2$. As the temperature increases, $\Delta I^*$ monotonically decreases and eventually disappears around $T_{\rm BKT}$ = 8 K where the voltage jumps become unclear (Figs. E1(a)-(b)). The temperature dependence is consistent with our scenario that the current-driven dynamics of dissociated vortex pairs is relevant to the SDE.

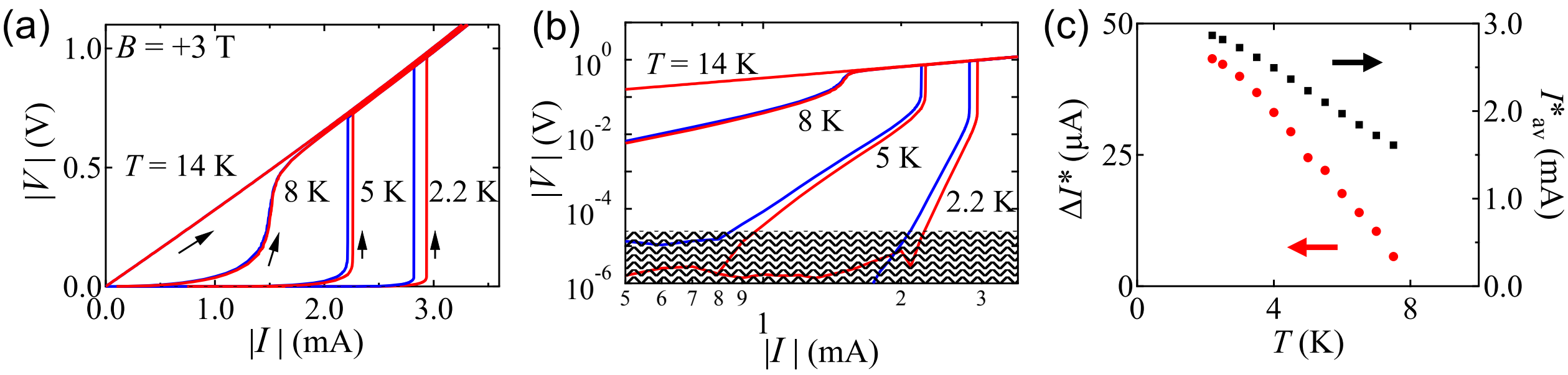

**FIG. E1. Temperature dependence of |*I*|-|*V*| characteristics**. (a) |*I*|-|*V*| curves at $T = 2.2$, 5.0, 8.0, and 14.0 K under the in-plane magnetic field of $B = +3$ T. The curves for $I > 0$ ($I < 0$) are shown in red (blue). (b) The logarithmic scale plot of the same data as in (a). The shadowed area lies below the noise floor. (c) Temperature dependence of $\Delta I^*$ (red) and $\Delta I^*_{\rm av} = \frac{|I^*_+|+|I^*_-|}{2}$ (black).

# Supplemental Material:

# Two-dimensional superconducting diode effect in topological insulator/superconductor heterostructure

Soma Nagahama[1], Yuki Sato[2], Minoru Kawamura[2], Ilya Belopolski[2], Ryutaro Yoshimi[3], Atsushi Tsukazaki[1,4], Naoya Kanazawa[5], Kei S. Takahashi[2], Masashi Kawasaki[1,2], and Yoshinori Tokura[1,2,6]

[1]Department of Applied Physics and Quantum-Phase Electronics Center (QPEC), The University of Tokyo, Tokyo 113-8656, Japan

[2]RIKEN Center for Emergent Matter Science (CEMS), Wako 351-0198, Japan

[3]Department of Advanced Materials Science, The University of Tokyo, Kashiwa 277-8561, Japan

[4]Institute for Materials Research (IMR),Tohoku University, Sendai 980-8577, Japan

[5]Institute of Industrial Science, The University of Tokyo, Tokyo 153-8505, Japan

[6]Tokyo College, The University of Tokyo, Tokyo 113-8656, Japan

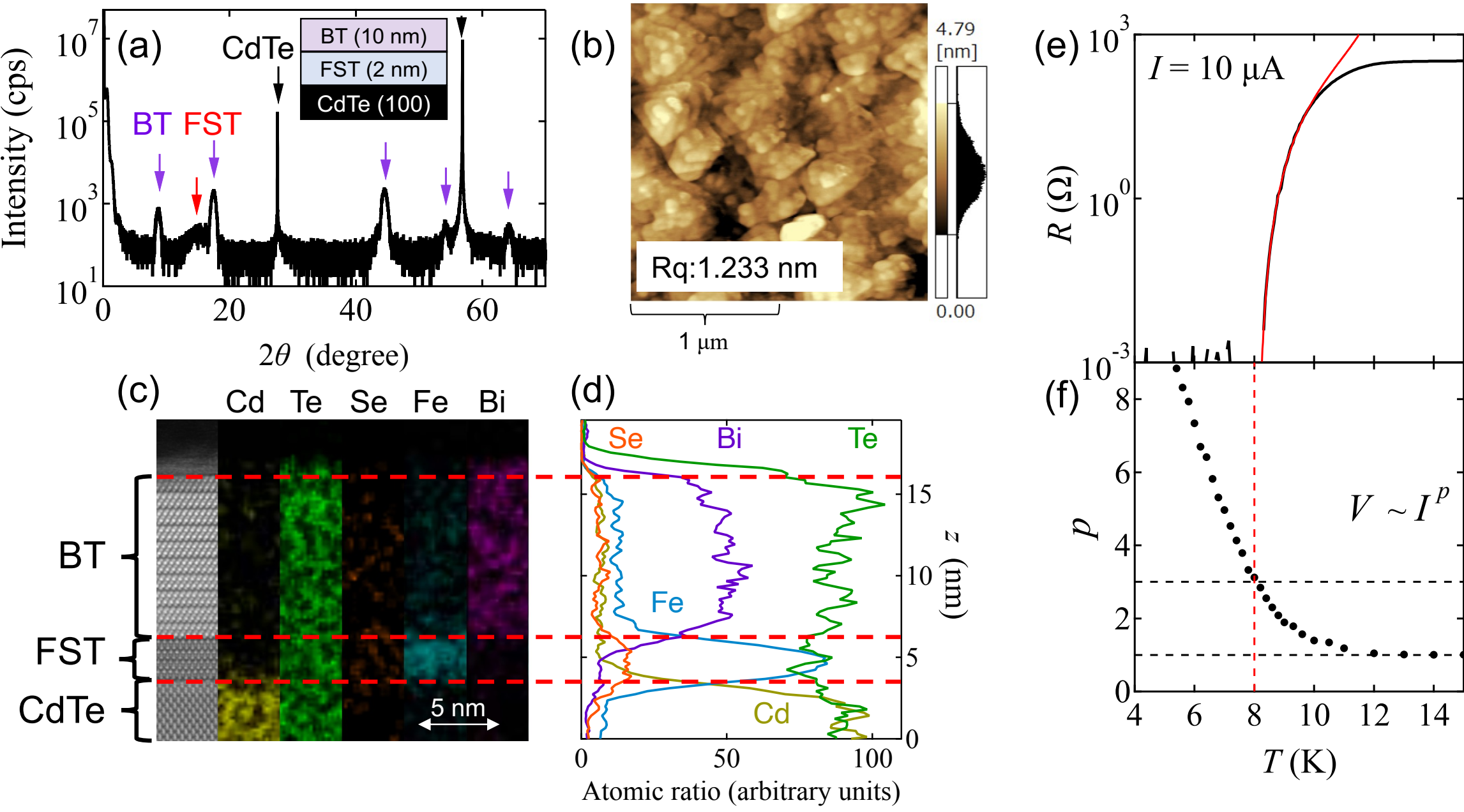

**FIG. S1. Characteristics of $Bi_2Te_3$/$FeSe_{0.1}Te_{0.9}$ (BT/FST) heterostructures**. (a) X-ray diffraction $2\theta$-$\omega$ scan data on a BT/FST heterostructure. Purple, red, and black arrows indicate the Bragg peak positions corresponding to $Bi_2Te_3$, $FeSe_{0.1}Te_{0.9}$, and CdTe, respectively. (b) Atomic force microscope image of a BT/FST heterostructure. Rq denotes the standard deviation of the height profile. (c) Cross-sectional high-angle annular dark-field scanning transmission electron microscopy image (left column) and elemental mappings for Cd, Te, Se, Fe, and Bi by energy dispersive x-ray spectroscopy (EDX). (d) Depth profile of the atomic ratio obtained by the EDX data in (c). Red dashed lines indicate the boundaries between each layer. (e) (Black line) Temperature dependence of the resistance plotted on a logarithmic scale. The resistance is measured with the excitation current of $I = 10$ μA. The red curve represents the fit using the Halperin-Nelson formula. (f) Temperature dependence of the exponent $p$ in the power-law $I$-$V$ characteristics $V \sim I^p$. The gray horizontal dashed lines indicate $p = 1$ and 3 respectively, and the red vertical dashed line indicates $T_{\mathrm{BKT}} = 8.0$ K.

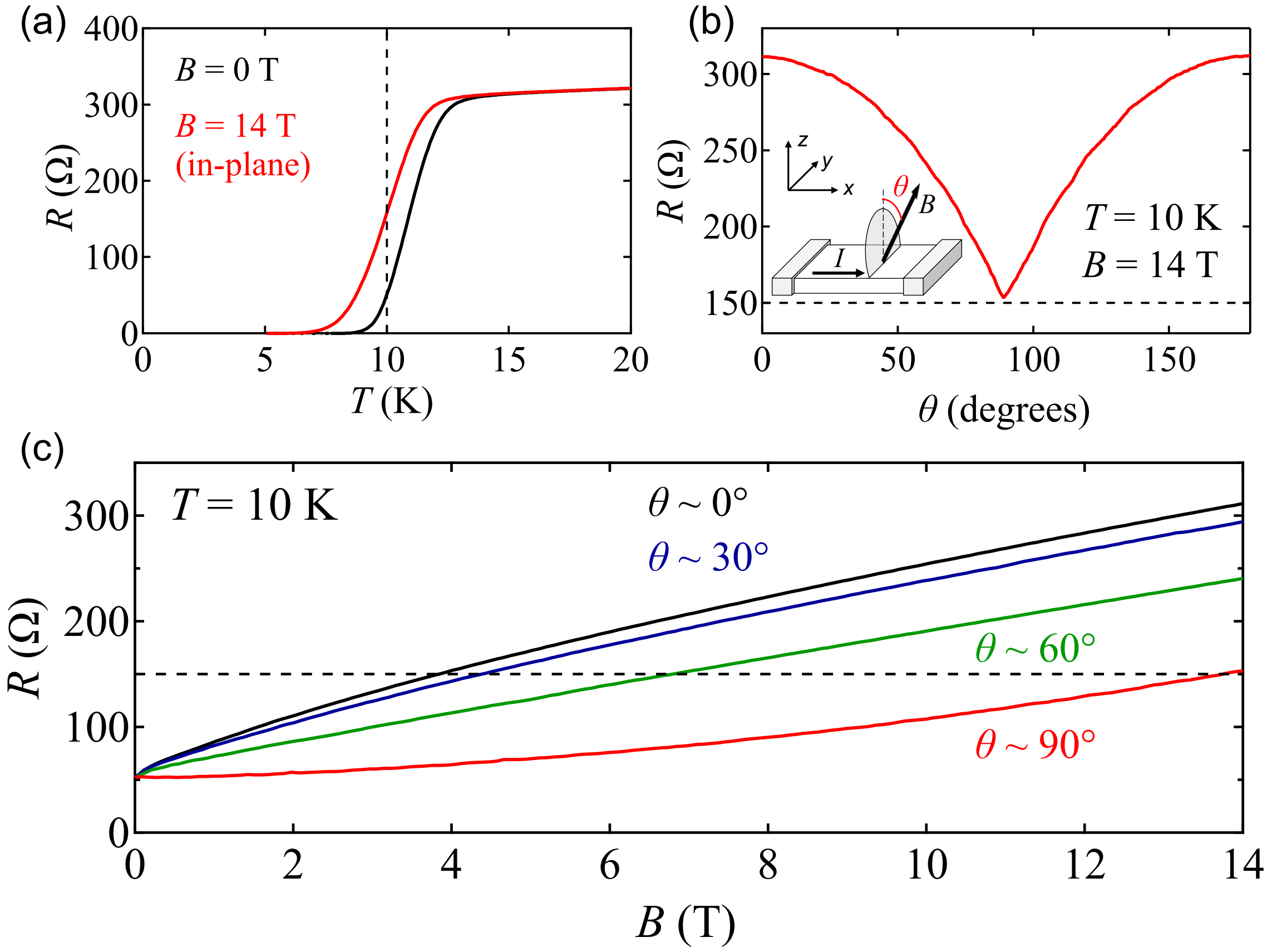

**FIG. S2. Magnetic field dependence of the resistance in the heterostructure.** (a) Temperature dependence of resistance (black) without magnetic field and (red) with magnetic field $B$ = 14 T. The vertical dashed line indicates $T$ = 10 K. (b) Angle dependence of resistance at $T$ = 10 K with magnetic field $B$ = 14 T. Dashed line indicates $R$ = 150 Ω. (c) Magnetic field dependence of the resistance with various angle of $\theta$.

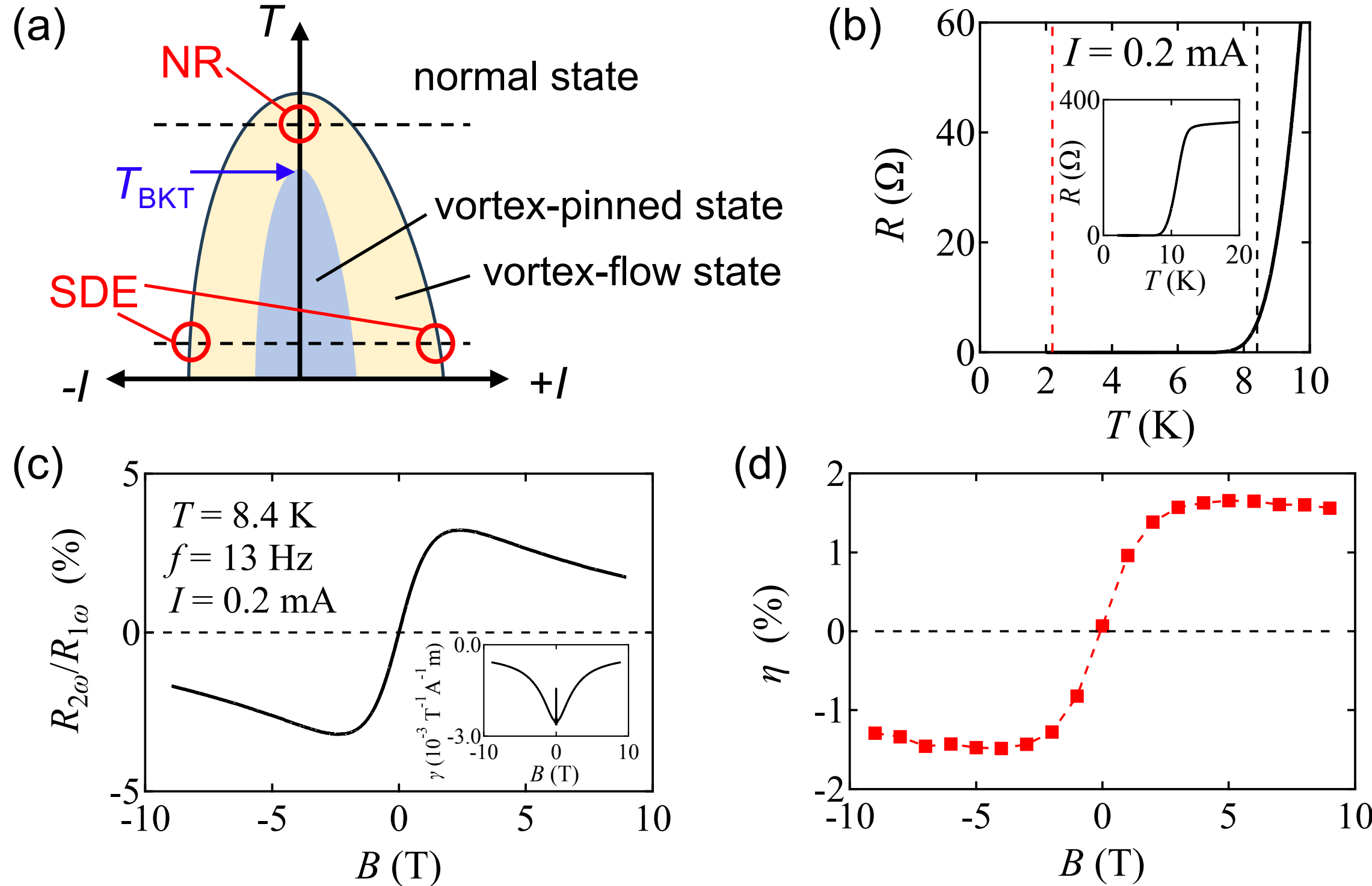

**FIG. S3. Relationships between the superconducting diode effect (SDE) and nonreciprocal resistance (NR) $R_{2\omega}$ above $T_{\mathrm{BKT}}$.** (a) Schematic temperature-current phase diagram of two-dimensional superconductivity. White, yellow, and blue regions indicate normal state, vortex-flow state, and vortex-pinned state, respectively. (b) Temperature dependence of the resistance of the BT/FST heterostructure. The inset displays the data in a wider temperature range. The red and black dashed lines respectively indicate the temperatures where SDE ($T$ = 2.2 K) and NR ($T$ = 8.4 K) measurements were conducted. (c) Magnetic field dependence of $R_{2\omega}/R_{1\omega}$ in the NR measurement. $R_{1\omega}$ is the primary resistance and $R_{2\omega}$ is out-of-phase component of the second harmonic resistance. They were measured by applying AC current $I = \sqrt{2}I_0 \sin 2\pi f t$ with $I_0 = 0.2$ mA and $f$ = 13 Hz. The measurement temperature was $T$ = 8.4 K. The inset shows magnetic field dependence of $\gamma = \sqrt{2}R_{2\omega}w/BR_0I_0$. $w$ = 100 μm is a sample width and $R_0$ = 5.47 Ω is resistance at $T$ = 8.4 K without magnetic field. The noise around $B$ = 0 T is arising from the 0/0 condition. (d) Magnetic field dependence of $\eta$ in the SDE measurement. The measurement temperature was $T$ = 2.2 K.

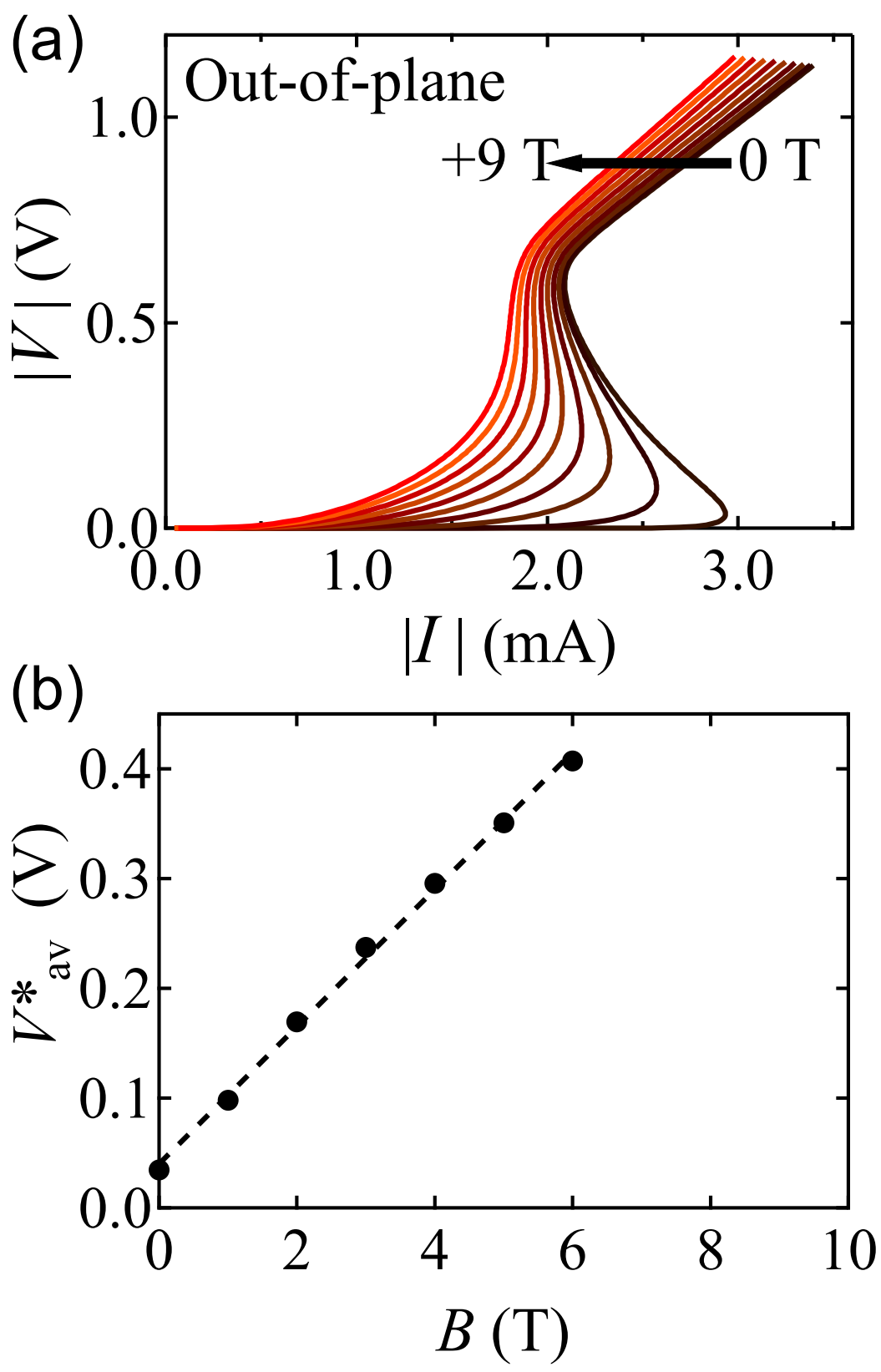

**FIG. S4. S-shape |*I*|-|*V*| characteristics under out-of-plane magnetic fields**. (a) |*I*|-|*V*| curves under out-of-plane magnetic fields obtained by the *V*-scan measurement. (b) Magnetic field dependence of $V_{\mathrm{av}}^{*}$ for the out-of-plane configuration. The dashed line is a linear fit.

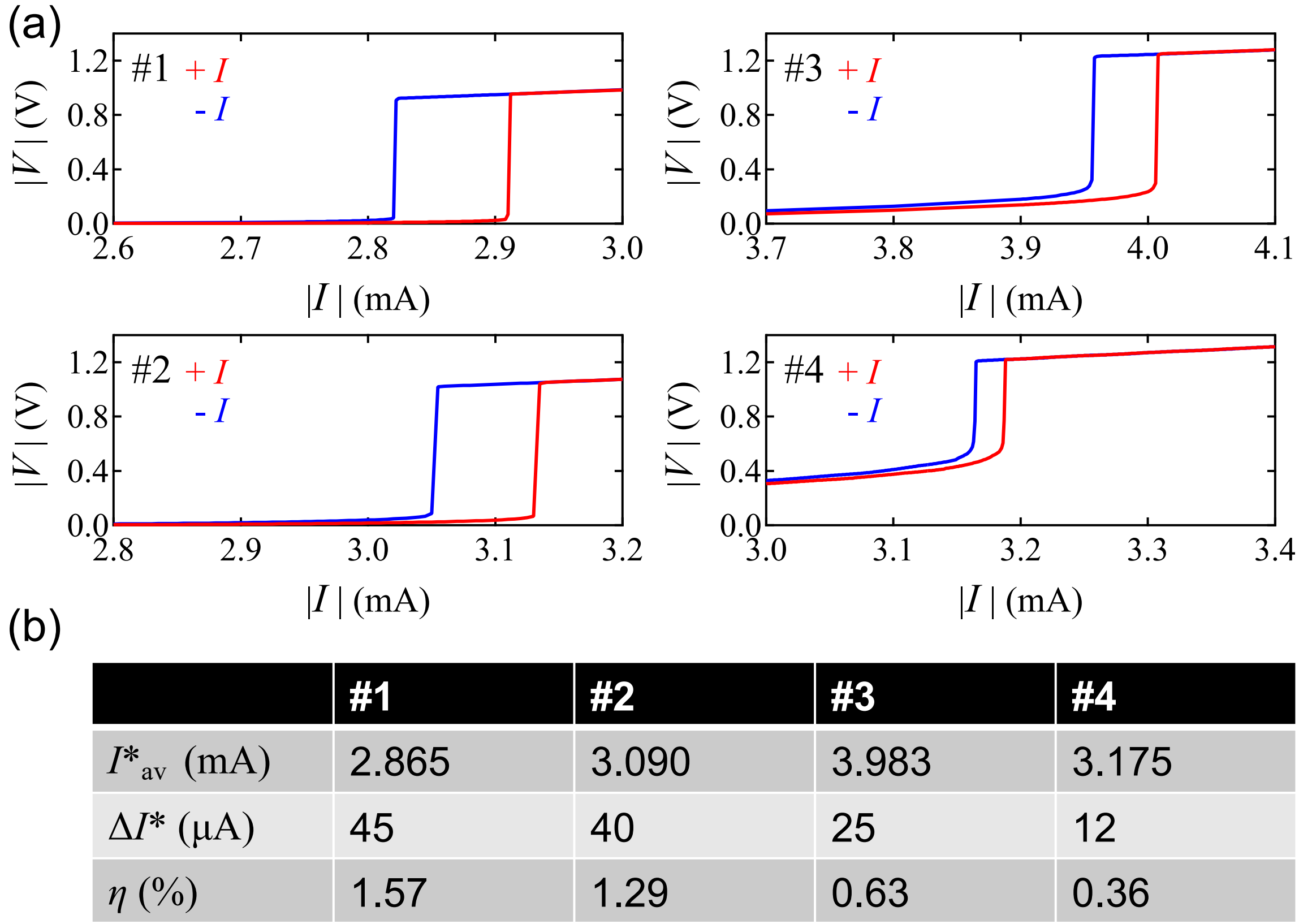

| | #1 | #2 | #3 | #4 |
|---|---|---|---|---|
| $I^*_{av}$ (mA) | 2.865 | 3.090 | 3.983 | 3.175 |
| $\Delta I^*$ (μA) | 45 | 40 | 25 | 12 |
| $\eta$ (%) | 1.57 | 1.29 | 0.63 | 0.36 |

**FIG. S5. Sample dependence of |*I*|-|*V*| characteristics**. (a) |*I*|-|*V*| curves under $B$ = 3 T for the four different samples. (b) $I^*_{\mathrm{av}} = \frac{|I^*_+|+|I^*_-|}{2}$, $\Delta I^*$ and $\eta = \Delta I^*/I^*_{\mathrm{av}}$ for the four samples. Sample #2 and #3 are used for Figs. 2(b) and 2(c) in the main text, respectively. Sample #1 is used for the other experiments.

## Supplementary Discussion 1 — Characterization of BT/FST heterostructure films

Figure S1(a) shows a $2\theta$-$\omega$ x-ray diffraction profile for a BT/FST heterostructure which is schematically illustrated in the inset. We observed clear Bragg peaks of BT, FST, and CdTe. The broadening of the peak of FST is attributed to the small thickness. We also performed atomic force microscopy on the BT/FST film (Fig. S1(b)). The standard deviation of the height profile, Rq, is found to be 1.233 nm, confirming that our films are reasonably flat. A cross-sectional high-angle annular dark-field scanning transmission electron microscopy (HAADF-STEM) image further reveals sharpness of the interfaces in the BT/FST film, as shown in Fig. S1(c). The elemental mappings for each element also confirm the successful growth of the desired heterostructure films without diffusion of atoms.

## Supplementary Discussion 2 — The 2D and 3D models in the angle dependence of $B_{c2}$

To confirm the 2D character in the superconducting properties in our samples, we studied the angle dependence of $B_{c2}$ by rotating the external magnetic field in the *yz*-plane (Fig. 1(f) in the main text). The angle dependence of $B_{c2}$ is one of the properties which reflects the dimensionality of a superconductor. In a 3D superconductor, the $B_{c2}$ with at an arbitrary angle $\theta$ is formulated as $\left(\frac{B_{c2}\sin\theta}{B_{c2,\parallel}}\right)^2 + \left(\frac{B_{c2}\cos\theta}{B_{c2,\perp}}\right)^2 = 1$, where $B_{c2,\parallel}$ and $B_{c2,\perp}$ are $B_{c2}$ under in-plane and out-of-plane magnetic fields, respectively. This model gives a smooth (or differentiable) curve over the entire range of $\theta$. On the other hand, in a 2D superconductor, $\theta$ dependence is formulated as $\left(\frac{B_{c2}\sin\theta}{B_{c2,\parallel}}\right)^2 + \left|\frac{B_{c2}\cos\theta}{B_{c2,\perp}}\right| = 1$ [24]. Notably, this model gives a cusp anomaly at $\theta = 90°$, in stark contrast to the 3D case. The $\theta$ dependence of $B_{c2}$ at $T = 10$ K is present in Fig. 1(f) of the main text. We defined $B_{c2}$ as the magnetic field where the resistance becomes $R = 150\ \Omega$, which is around half the normal resistance (Fig. S2). We measured $B_{c2}$ at $T$ = 10 K because the 2D model, which is derived using Ginzburg-Landau equation, is valid near the superconducting transition temperature [34]. In addition, we choose the particular temperature of 10 K so that the resistance values can be larger than 150 Ω for all $\theta$ with increasing magnetic field ($B$) up to 14 T, and hence we could measure the $\theta$ dependence of $B_{c2}$.

We plot the two models using experimentally determined values of $B_{c2,\parallel}$ and $B_{c2,\perp}$. The experimental data exhibits a cusp anomaly at $\theta$ = 90° and follows the 2D model well. This result supports the 2D character of the superconductivity in our samples.

## Supplementary Discussion 3 — Comparison between SDE and nonreciprocal resistance (NR) above BKT temperature

Both superconducting diode effect (SDE) observed at temperatures well below $T_{BKT}$ and nonreciprocal resistance (NR) observed at temperatures above $T_{BKT}$ represent nonreciprocal charge transport behavior. In our earlier study [23], we reported nonreciprocal resistance (NR) using the similar heterostructures to the present study. Here, we comment on the differences and similarities between the NR and the SDE. The NR can be described as $R = R_0[1 + \gamma \boldsymbol{z} \cdot (\boldsymbol{I}/w \times \boldsymbol{B})]$. Here, $R_0$ is the resistance under the zero magnetic field, $\boldsymbol{z}$ is the unit vector along $z$-axis, $w$ is the sample width, and $\gamma$ is the coefficient characterizing the magnitude of the nonreciprocity. NR is often detected using the AC lock-in experiment, where non-zero second harmonic resistance $R_{2\omega}$ = Im[$V_{2\omega}/I_0$] = -$\gamma R_0 B I_0/\sqrt{2}\,w$ appears as a response to the input current $I = \sqrt{2} I_0 \sin \omega t$. We performed NR measurements at $T$ = 8.4 K, which is slightly higher than $T_{BKT}$ (Fig. S3(b)). By applying AC current $I = \sqrt{2} I_0 \sin \omega t$ with $I_0$ = 0.2 mA, we measured the second harmonic resistance, given by $R_{2\omega}$ = -$\gamma R_0 B I_0/\sqrt{2} w$ which is an indicator of NR.

We emphasize that the SDE and the NR measurement were conducted under distinct experimental conditions, as schematically depicted in Fig. S3(a). The SDE was conducted at $T$ = 2.2 K well below $T_{BKT}$, whereas the NR measurement was performed at temperatures above $T_{BKT}$. In addition, the SDE involves substantially large currents ($I \sim 3$ mA, in the present study) where a transition to the normal state takes place, whereas the NR measurement is usually conducted with much smaller currents ($I_0$ = 0.2 mA).

Figure S3(c) shows the magnetic field dependence of $R_{2\omega}/R_{1\omega}$ It increases linearly with the magnetic field and saturates at around 3 T. The magnetic field dependence appears similar to that of $\eta$ derived from $\Delta I^*$, as shown in Fig. S3(d). This similarity suggests that the in-plane magnetic field could exert a similar influence on moving vortices in both cases, despite considerably different situations of vortex flow dynamics. In the NR measurement above $T_{BKT}$, thermally excited vortices are driven by the Lorentz force generated by the current. In contrast, during the SDE measurement, a relatively large current produces the Lorentz force strong enough to mobilize the nearly frozen vortices at the temperature.

## Supplementary Discussion 4 — Vortex-flow instability under out-of-plane magnetic fields

Figure S4(a) shows the *I*-*V* characteristics obtained by *V*-scan under various out-of-plane magnetic fields. With increasing magnetic field, the characteristic S-shape of the *I*-*V* curves becomes less pronounced. The voltage $V_{\mathrm{av}}^*$, which corresponds to the current peak in the S-shaped *I*-*V* curves, increases with increasing the magnetic field. As shown in Fig. S4(b), $V_{\mathrm{av}}^*$ changes linearly with the magnitude of the magnetic field, showing a stark contrast to the case when magnetic field is parallel to the film plane (Fig. 4(f) and 4(g) in main text).

In the vortex-flow regime, the generated voltage is proportional to both the vortex velocity $v$ and the vortex density $n$. Accordingly, the relationship between the voltage and vortex density is given by $V = \frac{h}{2e} ndv$, where $d$ = 100 μm is the distance between the two electrodes, $h$ is the Planck's constant, and $e$ is the elementary charge. When a magnetic field $B$ is applied normal to the film plane, the magnetically induced vortices are created with a density $n_{\mathrm{ind}} = \frac{2eB}{h}$. The nearly $B$-linear behavior of Fig. S4(b) is predominantly attributed to the $B$-dependence of $n_{\mathrm{ind}}$. Thus, from the slope of Fig. S4(b), the critical velocity at which the instability occurs can be estimated as $v = 6 \times 10^2$ m/s. The critical velocity appears to be much smaller than those in other systems [25,28,44], potentially reflecting the large effective mass in Te-rich FST [32].